\documentclass[useAMS,usenatbib]{mn2e}
\usepackage{graphicx}
\def\jcoph{J. Comp.\ Phys.}

\def\eg{{\it e.g.}}
\def\etal{{\it et al.}}
\def\etc{{\it etc.}}
\def\ie{{\it i.e.}}

\def\DF{{\small DF}}

\def\pmb#1{\setbox0=\hbox{$#1$}%
  \kern-0.25em\copy0\kern-\wd0
  \kern.05em\copy0\kern-\wd0
  \kern-0.025em\raise.0433em\box0}

\def\Ecirc{E_{\rm circ}}

\def\GCS{{\small GCS}}
\def\Hipp{{\small HIPPARCOS}}
\def\LSR{{\small LSR}}
\def\ILR{{\small ILR}}
\def\OLR{{\small OLR}}

\title[Detection of a Lindblad Resonance]{A Recent Lindblad Resonance
in the Solar Neighbourhood}
\author[J. A. Sellwood]{J. A. Sellwood$^{1}$\thanks{E-mail:
sellwood@physics.rutgers.edu} \\
$^{1}$Rutgers University, Department of Physics \& Astronomy, 136
Frelinghuysen Road, Piscataway, NJ 08854-8019, USA}

\begin{document}


\pagerange{\pageref{firstpage}--\pageref{lastpage}} \pubyear{2010}

\maketitle

\label{firstpage}

\begin{abstract}
Stars in the solar neighbourhood do not have a smooth distribution of
velocities.  Instead, the distribution of velocity components in the
Galactic plane manifests a great deal of kinematic substructure.  Here
I present an analysis in action-angle variables of the
Geneva-Copenhagen survey of $\sim 14\,000$ nearby F \& G dwarfs with
distances and full space motions.  I show that stars in the so-called
``Hyades stream'' have both angle and action variables characteristic
of their having been scattered at an inner Lindblad resonance of a
rotating disturbance potential.  This discovery seems to favour spiral
patterns as recurrent, short-lived instabilities.
\end{abstract}

\begin{keywords}
galaxies: evolution -- galaxies: kinematics and dynamics --
galaxies: spiral
\end{keywords}

\begin{figure*}
\includegraphics[width=.65\hsize,angle=270]{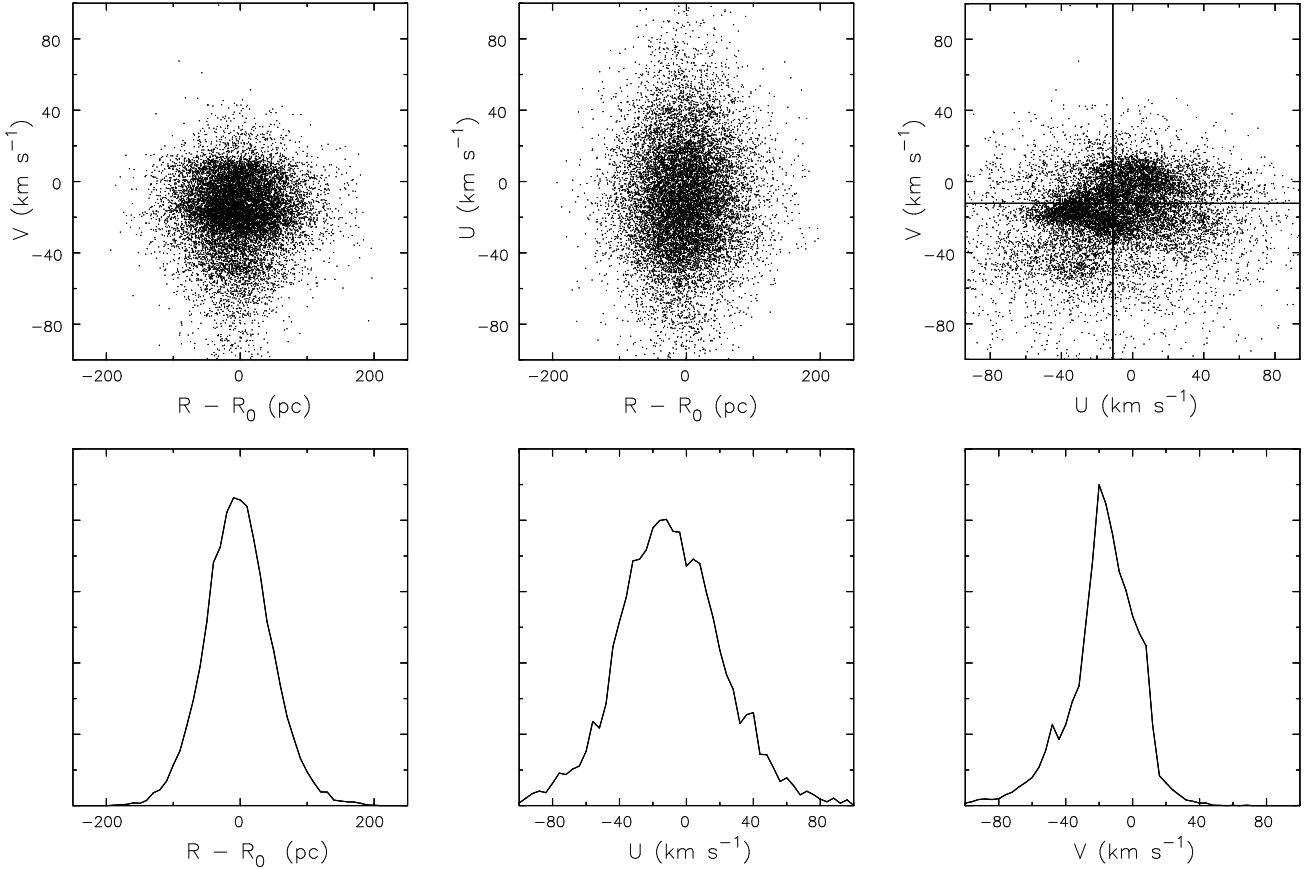}
\caption{The lower panels show the distributions of Galacto-centric
  distance $R - R_0$, and the velocity components $U$ (towards the
  Galactic centre) and $V$ (in the direction of Galactic rotation) for
  all Geneva-Copenhagen sample stars within 200~pc of the Sun.  The
  upper panels show three projections of the 3-D distribution; while
  there are no obvious correlations of either velocity component with
  Galacto-centric distance, the distribution in velocity space (top
  right) is highly non-uniform \citep{Dehn98,GCS}.  The lines in
  the top right panel mark the LSR estimated by \citet{SBD}.}
\label{dtbns}
\end{figure*}

\section{The Local Distribution of Stellar Velocities}
The \Hipp\ satellite \citep{ESA7} obtained accurate distances and
proper motions for $\sim 110\,000$ stars, but did not determine radial
velocities for any.  \cite{DB98} tried to construct an unbiased
sub-sample for their analyses of local stellar kinematics, and
estimated the missing radial velocities statistically \cite{Dehn98}.

The Geneva-Copenhagen survey of nearby stars \citep[hereafter
  \GCS,][]{GCS} was designed to supply the missing velocity component
for a homogeneous sample of $16\,682$ nearby F \& G dwarf stars.  This
gargantuan effort has resulted in a large sample of stars in the solar
neighbourhood with distances and full space motions, the latest
revision of which \citep{HNA9} forms the basis of this paper.

The greater part of the \GCS\ sample ($14\,139$ stars) have multiple
measurements of radial velocities to check for contamination by
binaries.  \cite{HNA9} used the improved distances from the reanalysis
of \Hipp\ data by \cite{vL07}, but substituted photometric distance
estimates for those stars with trigonometric uncertainties $> 13\%$.
Proper motions are from \Hipp\ and Tycho-2 \citep{Tycho}.  I do not
use the disputed age estimates \citep[\eg][]{Reid07,Holm07} in the
present paper.

This monumental survey is uniquely valuable, since it is free from
most of the selection biases that went into the full \Hipp\ sample.
Aside from a concentration of 112 stars in the Hyades cluster, their
distribution over the sky is remarkably uniform, with a slightly
higher density in the declination range south of $\delta = -26^\circ$.

The machine readable table produced by \cite{HNA9} includes the star
positions and the $U, V, \&\; W$ components of the star's motion
relative to the Sun in Galactic coordinates.\footnote{These Cartesian
  velocity components are oriented such that $U$ is towards the
  Galactic centre, $V$ is in the direction of Galactic rotation, and
  $W$ is towards the north Galactic pole.}  Of the $16\,682$ stars in
the table, 596 have no distance and 2536 have blank fields for the
$U$, $V$ \& $W$ velocities.  I also discard the 363 stars having
distances $>200\;$pc from the Sun and the 112 Hyades cluster stars
leaving $13\,045$ stars that I use in this analysis.

The median distance of these selected stars is 74~pc and the sample
within 40~pc of the Sun is believed to be near complete.  \cite{GCS}
do not supply individual uncertainties for each velocity, but assert
that they are believed accurate to 1.5~km~s$^{-1}$, with the greatest
contribution coming from distance uncertainties.  I find some evidence
to corroborate this claim in the analysis presented here.

While the 112 Hyades cluster stars may seem too few in number to
affect this analysis materially, I have omitted them because they are,
in fact, resonant stars and would give a spurious boost to the
significance of the main result (see \S\ref{restrs}).

\cite{Fama05} present a further sample of northern K and M giants with
distances and full space motions, but rather few are close enough to
have acceptably small velocity uncertainties.  I have therefore not
included their sample of giants in this analysis.

\begin{figure}
\includegraphics[width=\hsize]{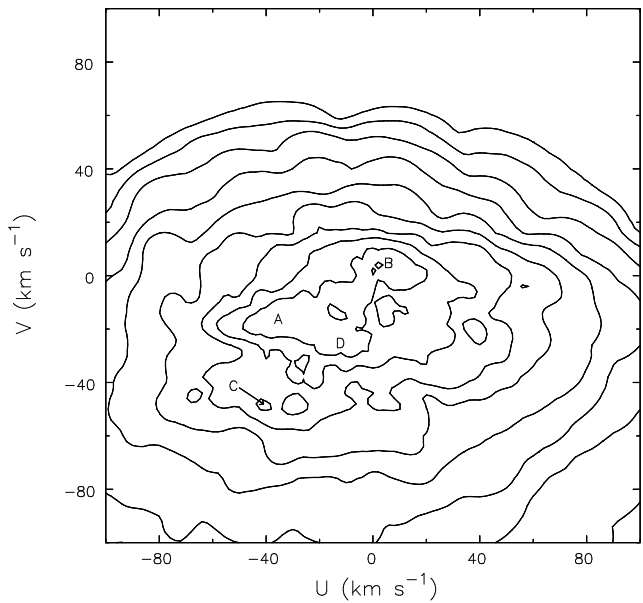}
\caption{Logarithmically spaced contours of the surface density of
  stars projected in the $U-V$-plane, as shown in Fig.~\ref{dtbns}.
  The letters indicate the approximate locations of the principal
  ``star streams'': A -- Hyades, B -- Sirius, C -- Hercules, and D --
  Pleiades.}
\label{uvcont}
\end{figure}

\subsection{Phase-space Distribution}
Fig.~\ref{dtbns} confirms the salient features of Dehnen's (1998)
results, which were based on stars lacking the radial velocity, and
which had residual concerns over selection biases.  The ``startling''
aspect \citep{BHR9} is the amount of substructure in the $(U,V)$
plane: the distribution function (\DF) of nearby stars appears to be
far from simple.  There is little in the way of an underlying smooth
component and the entire stellar distribution is broken into several
streams \citep{BHR9}.  The principal streams that enter into the
discussion below are labelled in Fig.~\ref{uvcont}.

The features are too substantial to have simply arisen from groups of
stars that were born with similar kinematics \citep[\eg][]{Egge96}, as
confirmed in detailed studies \citep{Fama07,Bens07,BH09}.  It is clear
that the entire \DF\ has been sculptured by dynamical processes.
While \cite{Helm06} suggest that some of the structure could be the
relics of satellite accretion, most work has focused on the effects of
the bar and spiral arms.

\cite{Kaln91} had previously suggested from entirely different data,
that the \OLR\ of the bar in the Milky Way might be close to the solar
circle.  After the release of data from \Hipp, \cite{Dehn00} also
attributed the Hercules stream (feature C) to the \OLR\ of the bar.
\cite{dSWT}, on the other hand reproduce distributions of stars with a
similar degree of substructure by invoking a succession of short-lived
spiral transients.  Other models that include both bars and spirals
are presented by \cite{Quil03}, \cite{Chak07} and \cite{Anto09}.

\cite{QM05} associate the Hyades stream feature with 4-fold symmetric
orbits in the potential of a long-lived 2-arm spiral.  Such orbits
arise near the ultra-harmonic (4:1) resonance of a finite amplitude
perturbation, where stars make four radial oscillations in a single
turn about the galaxy in a frame that rotates with the potential.

In all these studies, the spiral waves are put in ``by hand'', as
externally applied perturbations with properties of their choosing.
The supposed spiral waves should, of course, have arisen from the
stars themselves as self-consistent collective disturbances, and
therefore the properties of the waves assumed by these authors may not
be completely realistic.  In particular, their assumed shapes and
time-dependence may bias crucial resonant interactions with the stars,
which are responsible for the substructure they seek to explain.

\section{Distribution in Action Space}
However, the distribution in velocity space, which these various
authors attempt to mimic, is not the best projection of phase space to
reveal the origin of the structure.  A \DF\ that is in equilibrium can
be a function of the integrals only (Jeans theorem).  The classical
integrals for the in-plane motion in an axisymmetric disc are $E$ \&
$L_z$, but actions \cite[\eg][hereafter BT08]{BT08} are an alternative
set of integrals that have a number of advantages.

In a fully phase-mixed \DF, stars of a given set of actions are
uniformly distributed in the conjugate angle variables in a manner
that could produce a complicated structure when observed just in
velocity space.  Even if the \DF\ is not completely phase mixed, so
that Jeans theorem does not apply, its structure in action space is
unaffected by phase mixing, and therefore should be easier to
understand.  Furthermore, resonant interactions with the spirals, or
with the bar, are best viewed in integral space, since the only
lasting changes are to the integrals \citep{LBK}.  In this work,
therefore, I estimate action and angle variables from the observed
coordinates and spatial motions of the \GCS\ stars, and examine the
stellar distribution in the space of these variables.

For motion in a spherical potential, or the symmetry plane of an
axisymmetric potential $\Phi(R,z)$, the radial and azimuthal actions
are (BT08, p221)
\begin{equation}
J_R = {1 \over 2\pi} \oint \dot R dr, \quad\hbox{and}\quad
J_\phi \equiv L_z.
\label{actions}
\end{equation}
Here $\dot R$ is the radial speed of the star, the integral is taken
around a full radial period, and $L_z$ is the specific angular
momentum of the star about the Galactic centre.

The angles, $w_R$ \& $w_\phi$, conjugate to these actions both
increase at uniform rates defined by $\dot w_R \equiv \Omega_R \equiv
2\pi/T_R$ and $\dot w_\phi \equiv \Omega_\phi \equiv 2\pi/T_\phi$,
where the radial and azimuthal periods of the orbit are respectively
defined in equations (3.17) \& (3.19) of BT08.  Note in the limit of
nearly circular orbits, we have $\Omega_\phi \rightarrow \Omega_c$,
$\Omega_R \rightarrow \kappa$, and $J_R \rightarrow {1\over2}\kappa
a^2$, where $\Omega_c$, $\kappa$, and $a$ are respectively the angular
frequency of circular motion, the epicycle frequency, and the radius
of the star's epicycle.

I will here assume that the vertical motion of the disk stars is
decoupled from their in-plane motion, or equivalently that
the vertical action, $J_z$, is adiabatically invariant.  This is
justified because the vertical oscillation frequency, $\nu$, of stars
that do not climb to great heights above the disk plane is rapid
compared with the motion of the star in the plane.  This assumption
also requires that there are no resonances between the vertical motion
and rotating non-axisymmetric structures.  For disk stars whose
departures from circular motion are not too large, the variation in
the vertical frequency around the star's orbit can be neglected, and
the vertical action may be approximated as
\begin{equation}
J_z = {E_z \over \nu}, \quad\hbox{where}\quad E_z \simeq
{\textstyle{1\over2}}(z^2\nu^2 + v_z^2).
\label{evert}
\end{equation}
The approximate form for $E_z$ (BT08, eq.~3.86) holds for stars that
do not climb to great heights above the mid-plane.  Roughly two thirds
of the sample remain within 200~pc of the mid-plane, while just 5\%
can reach $z$-heights $\ga 500\;$pc.

Aside from a well-known tendency for older stars to have larger $J_z$
(or $E_z$), I have not found the distribution of $J_z$ values
revealing, and most distributions discussed in this paper are
integrated over all $J_z$.  I mention the effect of eliminating stars
with large $J_z$ in \S\ref{tests}.

\subsection{Evaluation of Action-Angle Variables}
Evaluation of the in-plane action and angle variables for the \GCS\
stars is impossible without knowledge of the mid-plane potential,
$\Phi(R,0)$, which is poorly known for the Milky Way.  Since the
rotation curve for the Galaxy in the neighbourhood of the Sun may not
be far from flat locally, I adopt the mid-plane potential
\begin{equation}
\Phi(R,0) = V_0^2\ln\left({R \over R_0}\right),
\end{equation}
with the $R_0$ being the radius of the solar circle, and $V_0$ being
the orbital speed of the local standard of rest (\LSR).  I correct for
solar motion by adding the peculiar velocity of the Sun $(11.1, 12.24,
7.25)\;$km~s$^{-1}$ \citep{SBD} to the tabulated $(U,V,W)$ components.
I note below (\S4) that the results in this paper are insensitive to
this choice.

In order to compute the four action-angle variables,
$(J_R,J_\phi,w_R,w_\phi)$, I integrate the orbit of every star in the
\GCS\ sample in this potential from its observed position.  I
normalize both actions by dividing them by the specific angular
momentum of the \LSR, $L_{z,0} = R_0V_0$; my adopted values of these
constants ($R_0 = 8\;$kpc and $V_0 = 220\;$km~s$^{-1}$) determine only
the scaling of the actions and other choices would not alter the
conclusions.  I choose $w_R=0$ at the apocentre of a star's epicycle
and $w_\phi=0$ at the azimuthal location of the Sun.

This procedure implies two major assumptions.  First, the Milky Way
rotation curve is locally flat; adjustment for a different potential
would merely distort the distribution of stars in the space of these
variables as points on a stretched rubber sheet.  Second, I have
assumed an axisymmetric potential, whereas the Milky Way certainly has
at least mild non-axisymmetry: the large majority of stars in the
\GCS\ sample have orbits that do not take them more than $\sim 2\;$kpc
from the orbit of the Sun, and most stay much closer.  This range of
distances is large enough for them to experience some non-axisymmetric
changes in the gravitational potential.  The instantaneous values of
these integrals at the solar azimuth must differ in a periodic manner
from those for similar stars at other azimuths, but the ordering of
stars in $J_\phi$ and $J_R$ should not be affected, since the \LSR\
also follows the same non-axisymmetric potential.  Thus, neither
assumption affects the topology of the distribution of stars in these
variables, and no features in the distribution could be erased or
appear if they were corrected.  Therefore, neither these assumptions,
nor uncertainties in the Galactic constants, $R_0$ \& $V_0$,
compromises the principal scientific finding of this paper.

\begin{figure}
\includegraphics[width=\hsize]{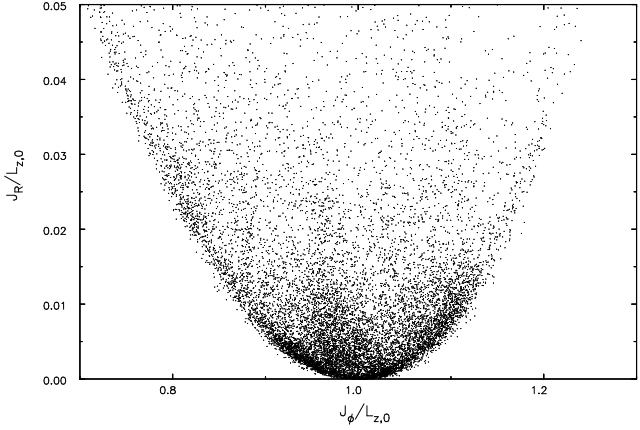}
\caption{The distribution of \GCS\ stars in the space of the two
  actions $J_R/L_{z,o}$ and $J_{\phi}/L_{z,0}$.  The approximately
  parabolic lower boundary reflects the selection of stars from the
  solar vicinity.  Aside from the general decrease in density towards
  large $J_R$ and a skew to lower $J_\phi$ caused largely by the
  asymmetric drift, the \DF\ in this projection also shows significant
  substructure.}
\label{actplt}
\end{figure}

\begin{figure}
\includegraphics[width=\hsize]{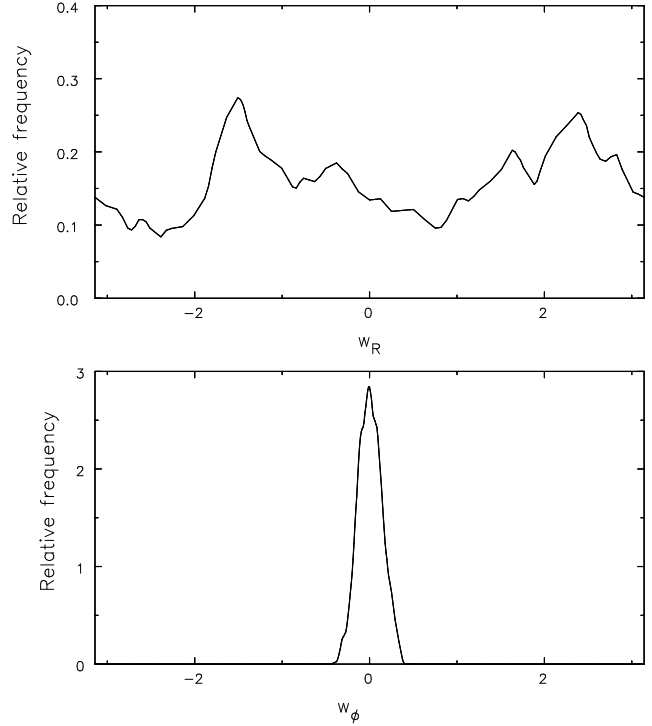}
\caption{The upper panel shows the distribution of angular phases,
  $w_R$, while the lower panel shows the distribution of azimuthal
  phases, $w_\phi$, for the \GCS\ stars.}
\label{angplt}
\end{figure}

\subsection{Results}
Fig.~\ref{actplt} shows the distribution of \GCS\ stars in action
space, excluding the small fraction (994 stars) with $J_R/L_{z,0} >
0.05$ (\ie\ with epicycle radii $a \ga 2\;$kpc).  The near-parabolic
lower boundary of the distribution reflects the fact that all the
stars in this sample are currently within 200~pc of the Sun, and
therefore must have ever more eccentric orbits (larger $J_R$) as their
guiding centre radii, or $J_\phi$, differ increasingly from that of
the Sun.

The general decrease in density with increasing $J_R$ is expected in
any reasonable stellar distribution, and the asymmetry between left
and right results from the fact that the star density increases
towards the Galactic centre; more stars visit the solar neighbourhood
from the interior, which gives rise to the asymmetric drift.  It is
clear from this Figure that the distribution in the space of these
integrals also has substructure; the most notable feature is a
pronounced overdensity of stars rising with slightly negative slope
from $J_\phi \la L_{z,0}$.

Fig.~\ref{angplt} shows the distributions of angle variables for the
same stars.  We should expect fewer stars in the upper panel near
$w_R=\pm\pi$ and near $w_R=0$, because such stars are visiting the
solar neighbourhood at respectively the peri- and apo-centres of their
orbits.  The fact that the peaks are not symmetrically placed near
$w_R = \pm\pi/2$ is interesting, as discussed below.  The azimuthal
phase distribution has a single strong peak near $w_\phi=0$ because
the stars are all local to the solar neighbourhood.  The width of this
peak, which has a half-width at half-maximum of $\sim \pm11\degr$,
reflects the spread in Galactic azimuths of the guiding centres for
stars currently passing by the Sun.

\begin{figure}
\includegraphics[width=\hsize]{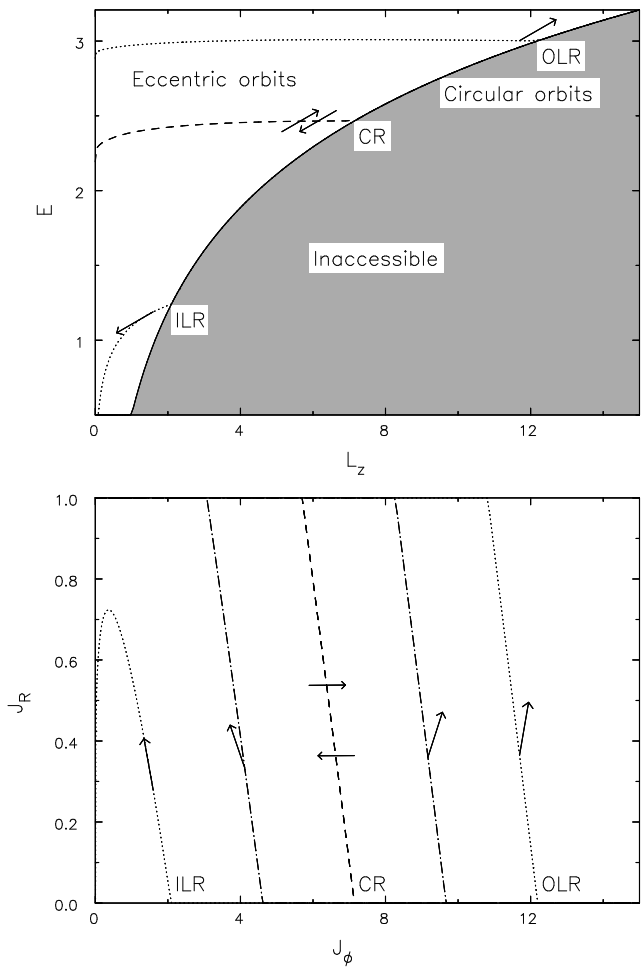}
\caption{The upper panel shows the Lindblad diagram for a simple disk
  galaxy model.  Circular orbits lie along the full-drawn curve and
  eccentric orbits fill the region above it.  Angular momentum and
  energy exchanges between a wave and stars move them along lines of
  slope $\Omega_p$ as shown.  The dotted and dashed lines show the
  loci of resonances for an $m=2$ perturbation of arbitrary pattern
  speed.  The lower panel plots the same resonance lines and
  scattering vectors in action space, where $J_\phi=L_z$ and circular
  orbits lie on the axis $J_R=0$.  The dot-dashed lines in the lower
  panel show the Lindblad resonances for an $m=4$ disturbance having
  the same pattern speed.}
\label{lindblad}
\end{figure}

\section{Exchanges at Resonances}
The upper panel of Fig.~\ref{lindblad} shows the Lindblad diagram for
a Mestel disk with an infinitesimal perturbation of constant pattern
speed $\Omega_p=0.14$, chosen arbitrarily.  The full-drawn curve marks
the locus of circular orbits in this $(L_z,E)$ plane; no star can lie
below this curve, but bound stars with $E>\Ecirc$ move on eccentric
orbits in this potential.

When the potential includes an $m$-fold rotational symmetric
perturbation with pattern speed $\Omega_p$, stars orbiting in the disc
encounter wave crests at the Doppler-shifted frequency
$m|\Omega_p-\Omega_\phi|$.  Resonances arise when
\begin{equation}
m(\Omega_p-\Omega_\phi) = l\Omega_R,
\end{equation}
with $l=0$ at corotation, and $l=\pm1$ at the Lindblad resonances
where the guiding center respectively overtakes, or is overtaken by,
the wave at the star's unforced radial frequency.\footnote{Note
  $|l|=1/2$ at ultraharmonic resonances.}  The loci of these three
principal resonances for arbitrarily eccentric orbits are marked by
the broken curves in this Figure, which intersect the circular orbit
curve where the more familiar epicyclic definitions of resonances
apply.  These lines are drawn for a bi-symmetric ($m=2$) disturbance;
the Lindblad resonances would be closer to corotation for $m>2$.

Stars moving in a non-axisymmetric potential that rotates at a steady
rate conserve neither their specific energy, $E$, nor their specific
angular momentum, $L_z$.  But the combination
\begin{equation}
I_{\rm J} \equiv E - \Omega_p L_z,
\end{equation}
known as Jacobi's invariant (BT08, eq. 3.112), is conserved.
Therefore, when a star changes its angular momentum by an amount
$\Delta L_z$ caused by the rotating potential disturbance, it also
changes its energy by the amount
\begin{equation}
\Delta E = \Omega_p \Delta L_z.
\end{equation}
Thus changes to these classical integrals caused by interactions with
a steadily rotating disturbance have slope $\Omega_p$ in the Lindblad
diagram.

\cite{LBK} showed that a lasting change to the angular momentum of a
star can occur only at resonances.  As the slope of the circular orbit
curve at corotation is $\Omega_p$, stars that exchange energy and
angular momentum there do not move away from that curve, to first
order, enabling exchanges to occur without heating, as described by
\cite{SB02}.  The other vectors show that exchanges at the Lindblad
resonances, on the other hand, cause stars to move onto more eccentric
orbits (farther from the circular orbit curve) when angular momentum
is redistributed outwards.  This is the root cause of disk heating by
spirals.

The lower panel of Fig.~\ref{lindblad} shows in action space the same
resonance lines and scattering vectors as the upper panel shows for
the space of the classical integrals.  Exchanges at corotation cause
horizontal shifts, implying no increase in $J_R$, while gains in
$J_\phi$ at the outer Lindblad resonance, hereafter \OLR, and losses
at the inner Lindblad resonance (\ILR) both cause increases in $J_R$.
The labelled resonance lines are drawn for $m=2$ and an arbitrarily
chosen value of $\Omega_p$; the diagram for other adopted pattern
speeds in this scale-free potential would have a similar appearance,
but with the $J_\phi$-axis rescaled.  The dot-dash lines in the lower
panel mark the Lindblad resonances for an $m=4$ wave of the same
pattern speed, with the scattering vectors again shown.

Notice that the direction of the scattering vector closely follows the
resonant locus for $m=2$ (dotted curve) only at the \ILR.  Thus, when
stars are scattered at this resonance, they stay on resonance as they
gain random energy, allowing very strong scatterings to occur.  The
opposite case arises at the \OLR, where the star is moved off
resonance by a small gain of angular momentum.  The alignment is
closest for $m=2$ disturbances in potentials that yield flat rotation
curves, but scattering is always stronger at the \ILR\ than at the
\OLR\ for $m \geq 2$ and in more general potentials.

All three resonance lines have similar, though not identical, slopes
in the action plot and for all $m\geq2$.  Thus resonant stars will lie
along a line in action space that has approximately the same negative
slope for any resonance.  However, we can identify the resonance at
which they were scattered or trapped by examining the angles, since
the resonance will have reset the values of the phase angles for
trapped stars to have the relation
\begin{equation}
m w_\phi + l w_R = \hbox{constant},
\label{wphase}
\end{equation}
where $l$ and $m$ were defined above.  Both angles increase with time
but the equation continues to hold at any fixed time because resonant
stars have $m \dot w_\phi + l \dot w_R = m \Omega_p$.  Thus the value
of the ``constant'' in eq.~(\ref{wphase}) varies as
$m\Omega_p(t-t_0)$, but from an unknown value at some earlier $t_0$
that depends upon the phases of the stars relative to the
perturbation.  We will use eq.~(\ref{wphase}) in \S~\ref{phases}.

\section{Identification of Resonances}
While the principal feature in Fig.~\ref{actplt} is clear to the eye,
I need a method to determine its statistical significance and to
search for possible weaker features.  This requires a statistic that
contrasts the data with a comparable featureless model.  Ideally, a
suitable featureless model would have a simple distribution function
but the velocity distributions shown in Fig.~\ref{dtbns} are not
simple; the $U$-distribution is not Gaussian (it has a large
kurtosis), while the $V$-distribution is complex and skew.  Thus no
simple function of the integrals could yield a \DF\ that would be an
approximate match to the density of \GCS\ stars in action space.

\begin{figure}
\includegraphics[width=\hsize]{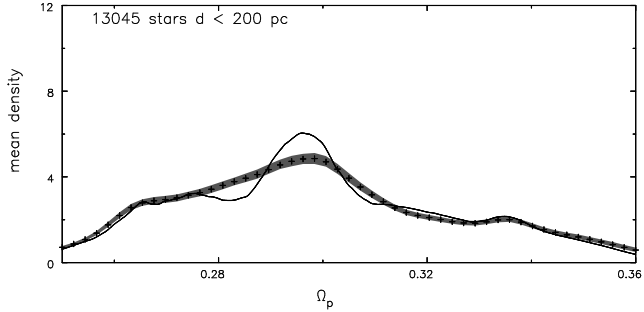}
\caption{The solid line shows mean density of \GCS\ stars in the
  $(J_\phi,J_R)$ plane along scattering trajectories from circular
  orbits, as a function of the pattern speed of the perturbation.  The
  shaded region shows the 99\% confidence limits from similar size
  star samples with randomly selected Galacto-centric distance and
  velocity components.}
\label{linint}
\end{figure}

\subsection{Statistical Significance}
I therefore adopt a bootstrap approach in which I resample the actual
distributions of $R$, $U$ \& $V$ to generate comparison distributions
of pseudo stars.  Selecting all three coordinates independently
destroys the correlations that give rise to any features in integral
space.  While such a procedure could be dangerous if the velocity
components have large and widely differing errors, the velocity and
distance errors in the present sample are all small with respect to
the ranges of the data.  Samples of pseudo data selected in this way
populate the $(J_\phi,J_R)$-plane in a very similar manner to the real
data, but with any small-scale structure smoothed away.

As a quantitative estimator of structure, I compute the mean surface
density of stars in $(J_\phi,J_R)$ space along resonance lines.  I
estimate the surface density of stars in the $(J_\phi,J_R)$-plane
using an adaptive smoothing kernel \citep{Silv86} of elliptical shape
with axis ratio 12:1 to compensate for the differing ranges of
$J_\phi$ and $J_R$.  To compute the mean density as a function of
$\Omega_p$, I integrate the estimated star density in this plane along
a resonance line for that selected $\Omega_p$.  The path of
integration is from where the resonance line intersects the lower
boundary of the stars to $J_R=0.05$; dividing by the length of this
line yields the mean density.

Figure~\ref{linint} compares the mean density along \ILR\ lines for a
range of $\Omega_p$ with 99\% confidence intervals (shaded) of the
same quantity selected from 2000 samples of scrambled data.  The
Figure indicates one highly significant feature with a pattern speed
$\Omega_p \sim 0.296 \pm 0.005$.  No other feature is nearly as
significant.  Repeating this analysis for trapping at corotation and
at the \OLR, again revealed a single feature of high significance.

In \S\ref{tests}, I report the results of tests for possible selection
biases in this sample of stars that might affect the significance of
the feature detected in Figure~\ref{linint}.

Figure~\ref{linint} assumed a bisymmetric spiral pattern, but it is
possible the resonance arose from a spiral having higher rotational
symmetry.  Repeating the analysis assuming $m=3$ \& 4 revealed maxima
that were no less significant from that for $m=2$ but having different
frequencies.  Thus not only do the action data not determine which
resonance is responsible, but they do not determine the rotational
symmetry of the pattern either.

\begin{figure}
\includegraphics[width=\hsize]{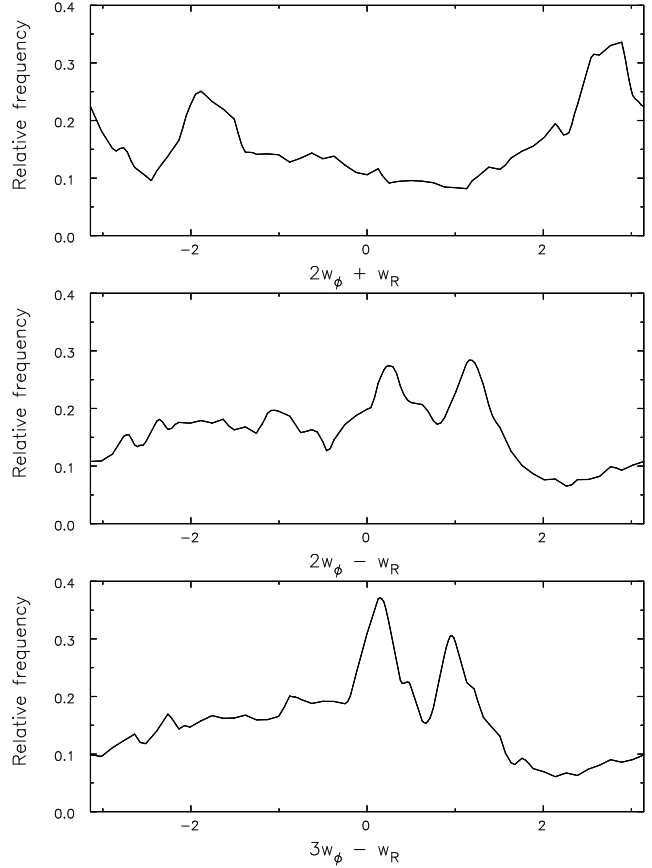}
\caption{The distribution of three linear combinations of phase angles
  as tests for correlations characteristic of resonances.}
\label{restest}
\end{figure}

\begin{figure}
\includegraphics[width=\hsize]{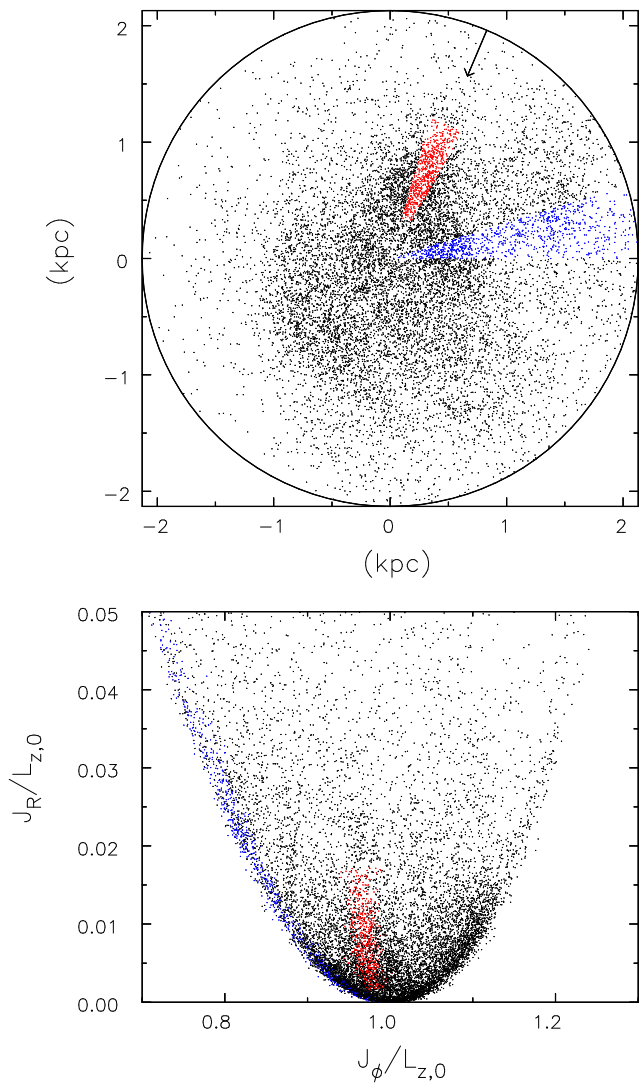}
\caption{The upper panel shows the distribution of \GCS\ stars in the
  space of the epicycle amplitude and spiral phase.  The radial
  coordinate is $a=(2J_R/\kappa)^{1/2}$, and the azimuthal coordinate
  is $2w_\phi-w_R$.  The lower panel is the same as shown in
  Fig.~\ref{actplt}, except it uses the colours assigned in the upper
  panel.}
\label{polar}
\end{figure}

\subsection{Phases}
\label{phases}
We can tell which resonance was responsible for the density excess in
the $(J_\phi,J_R)$-plane by examining the distribution of phase angle
variables for the stars.  The distribution of $w_R$
(Fig.~\ref{angplt}) has two peaks away from zero radians, while the
distribution of $w_\phi$ values is narrow because all the stars in the
sample are near the Sun.  Because of this selection effect, we should
expect the distribution of combinations of $mw_\phi \pm w_R$ to
resemble that of $\pm w_R$ alone, were the two angle variables
uncorrelated.

Fig.~\ref{restest} shows the distributions, modulo $2\pi$, of three
linear combinations of the two angle variables.  Our expectation is
borne out in the top panel, but not in the two lower panels.  The
change in shape of the distribution is quite dramatic, indicating
significant correlations that are characteristic of an \ILR\ for a
pattern.  The lower two panels assume patterns having two- or
three-fold rotational symmetry ($m=2$ or 3).  The peak near zero,
which becomes more significant as $m$ is increased, is a selection
effect, while the feature of most interest is the second peak away
from zero, which is still present for $m=4$ (not shown).  It is made
up of stars having negative $w_R$ values near the peak shown in
Fig.~\ref{angplt} that also have slightly negative values of $w_\phi$.

To investigate further, the upper panel of Fig.~\ref{polar} shows the
distribution of the \GCS\ stars in polar coordinates of the epicycle
radius $a=(2J_R/\kappa)^{1/2}$ and polar angle $2w_\phi-w_R$ as
appropriate for stars trapped at an \ILR\ of an $m=2$ pattern.  The
two peaks in the middle panel of Fig.~\ref{restest} indicate
concentrations of stars near $2w_\phi-w_R \sim 0.15$ and 1.17, or
9\degr\ and 67\degr.  I have coloured the stars in these
concentrations; all those having $2w_\phi-w_R = 67\degr \pm 7\degr$
and moderate radial action are coloured red, while those in the range
$2w_\phi-w_R = 9\degr \pm 9\degr$ are coloured blue.

I have then used this colouring to show, in the lower panel, the
locations of the same stars in the plot of actions, reproduced from
Fig.~\ref{actplt}.  The blue stars are concentrated along the boundary
of the lower plot, indicating that they are all close to the
apocentres of their orbits and the concentration at this phase,
discussed below, is probably of little dynamical significance.
However, the red stars coincide with the resonance peak identified
earlier.

The boundaries of the triangle in the upper panel of Fig.~\ref{polar}
enclose 639 stars.  These geometrically simple boundaries are not
intended to select all or only resonant stars, but it is likely that
most of the selected stars are resonant.  It is also likely that
additional resonant stars lie outside these somewhat arbitrary
boundaries.

If we assume a $m$-arm spiral, we should change the polar angle to
$mw_\phi-w_R$.  Polar plots for the cases $m=3$ \& 4 (not shown) are
very similar to the upper panel of Fig.~\ref{polar} except that the
``red'' stars are now concentrated at smaller angles, with almost
equal significance.  Furthermore, almost the {\it same set\/} of stars
form the local peak of the distribution of $mw_\phi-w_R$ for $m = 3$
\& 4 as for $m=2$, and they therefore again lie in the density excess
in the action plot (lower panel).

It is the correspondence between the stars of the selected phases that
also have values of their actions to place them in the scattering peak
that provides the most compelling evidence for an \ILR.  But once
again the evidence cannot discriminate whether the stars were trapped
by an $m=2$ wave, or one having higher rotational symmetry.

\subsection{Further Tests}
\label{tests}
I have conducted an extensive series of tests to check the result in
Fig.~\ref{linint}.  The main concern is that the apparently resonant
stars stand out because they share some other property unrelated to a
resonance.

Some 37\% of the \GCS\ stars are known or suspected binaries.  The
significance of the resonance peak is unaffected by including or
eliminating the stars flagged as possible or known binaries.

Increasing the kernel width used to estimate the density in the
$(J_\phi,J_R)$-plane (default $= 0.02$) smoothed the curve in
Fig.~\ref{linint} and lowered the maximum as expected.  But it also
reduced the ranges of the confidence intervals, so the feature
remained highly significant.

The principal source of uncertainty in the input data arises from the
uncertainty in the distance to each star, which generally rises with
distance to the star.  I therefore repeated the analysis using stars
lying within various distances from the Sun ($d<50\;$pc at the most
extreme), which had little effect on either the significance or the
position of the principal peak in the upper panel of
Fig.~\ref{linint}.

Performing the same analysis on stars separated into opposite
hemispheres on the sky, either in equatorial or Galactic coordinates,
did not affect the significance of the peak.  Since the spectroscopic
radial velocities could be more precise than those derived from the
proper motion, I tried separating the stars into samples within
$60^\circ$ of the Galactic poles from those lying closer to the plane.
The radial velocity contributes little to the $U \; \&\; V$ velocity
components for stars near the caps, but the scattering peak in this
case is just as significant as for the remaining stars.  I also tried
separate analyses for stars in quadrants centred on the Galactic
centre and anti-centre, from those in quadrants centred on the apex
and ant-apex of Galactic rotation, again finding little change.

Finally, I investigated the effects of dividing the sample using by
the energy of vertical motion, $E_z$ (eq.~\ref{evert}).  The resonance
peak is most significant among the stars in the lowest $\sim 66\%$ of
the vertical energy range, and much less significant in the other
33\%.  Such a dependence seems entirely reasonable, since stars with
larger vertical oscillations will be much less affected by in-plane
potential variations.

These various tests have proved reassuring.  The effects of selecting
subsamples of the stars are all entirely consistent with physical
expectations, and the likely influence of the small errors in the
quantities tabulated in the catalogue.

\subsection{Adjustments to the LSR}
The local standard of rest (\LSR), which is the speed of a particle on
a notional ``circular orbit'' passing through the Sun's position, may
not even be well-defined because the Galaxy is at least mildly
non-axisymmetric.  Furthermore, the Sun's speed relative to the
\LSR\ is hard to determine, and estimates by Binney and his
collaborators \citep{DB98,AB09,SBD} from the \Hipp\ data have changed
by several km~s$^{-1}$ at each revision.

As the values of the action-angle variables depend on the choice of
the \LSR, I have experimented with adopting values that differ from
the latest estimate \citep{SBD} by $\pm10$ and $\pm20\;$km~s$^{-1}$ in
both $U$ and $V$.  Shifting the $V$-component essentially shifts the
entire distribution in Fig.~\ref{actplt} to the left or right while
preserving most of the structure.  For larger revisions to the \LSR,
the upward rising tongue becomes rather broader, however.
Furthermore, the distribution of the angle variables is not changed
qualitatively, and I always obtain peaks in the distribution of $m
w_\phi - w_r$ for $m = 2$, 3 \& 4 that are located away from the
direction towards the Galactic centre.  Thus the main finding of this
paper is little affected by the precise choice of the \LSR.

\section{Discussion}
Fig.~\ref{polar} provides strong evidence that a small fraction ($\sim
5\%$) of the \GCS\ stars have been trapped at an \ILR\ of a rotating
disturbance.

\begin{table}
\caption{Estimated pattern speeds and corotation radii for the fitted
  disturbance.  The possible systematic errors in these quantities are
  much greater than the small statistical uncertainties.}
\label{table}
\begin{tabular}{@{}cccc}
Angular periodicity   & $\Omega_pR_0/V_0$  & $\Omega_p$ & $R_{CR}$ \\
   &  & km s$^{-1}$ kpc$^{-1}$ &  kpc \\
\hline
$m=2$ & $ 0.296$ & $ 8.1$ & 27 \\
$m=3$ & $ 0.534$ & $14.7$ & 15 \\
$m=4$ & $ 0.655$ & $18.0$ & 12 \\
\hline
\end{tabular}
\end{table}

\subsection{Angular Periodicity}
It is indeed surprising to find evidence for an \ILR\ near the Sun, as
it seems to imply a spiral pattern extending outwards from the
Galactic radius of the Sun's orbit.  However, none of the foregoing
analysis could discriminate whether the \ILR\ was that of a
bisymmetric pattern or one of higher rotational symmetry.

Assuming $R_0=8\;$kpc and $V_0=220\;$lm~s$^{-1}$, and using the
evidence that the scattering peak was caused by an \ILR, we may
determine the pattern speed of the disturbance from the frequency at
which the mean density in Fig.~\ref{linint} peaks.  Table~\ref{table}
gives, for each angular periodicity, the measured frequency of the
peak, and the implied pattern speed and corotation radius, also
assuming an exactly flat rotation curve.  (The statistical uncertainty
in the measured $\Omega_p$ is $\la2$\%, but systematic errors arising
from the various assumptions are likely to be much larger.)  While the
Milky Way rotation curve may not be exactly flat, it is clear that
corotation lies in the far outer disk for all of these patterns.

If the disturbance responsible for the \ILR\ is a spiral pattern (it
could be some other slowly rotating disturbance), one would expect
corotation to lie within the disk of the Milky Way, which probably
does not extend much beyond a Galacto-centric radius of 14~kpc
\citep[][but see also Carraro \etal\ 2010]{RCM92}.  This argument
would strongly disfavour an $m=2$ spiral, but the Milky Way disk may
be extensive enough to accommodate spirals of higher rotational
symmetry.

\cite{QM05} calculated the appearance of the $(U,V)$-plane in a simple
galaxy model that was perturbed by a large-amplitude, long-lived,
bi-symmetric spiral that rotated with various pattern speeds.  For a
pattern speed low enough to place the \ILR\ near the Sun, the
phase-space structure they calculated did not correspond well with
that observed, but higher pattern speeds that placed the Sun near the
ultraharmonic resonance of their assumed spiral created a feature
resembling the Hyades stream.  They showed that the orbits responsible
were 4:1 resonant periodic orbits in their adopted pattern.  Unlike in
their work, my analysis makes no assumption about the spatial form or
lifetime of the potential perturbation, but the fact that they
favoured 4:1 periodic orbits as the cause of the Hyades feature may be
hinting that an $m=4$ spiral perturbation is favoured.

\begin{figure}
\includegraphics[width=\hsize]{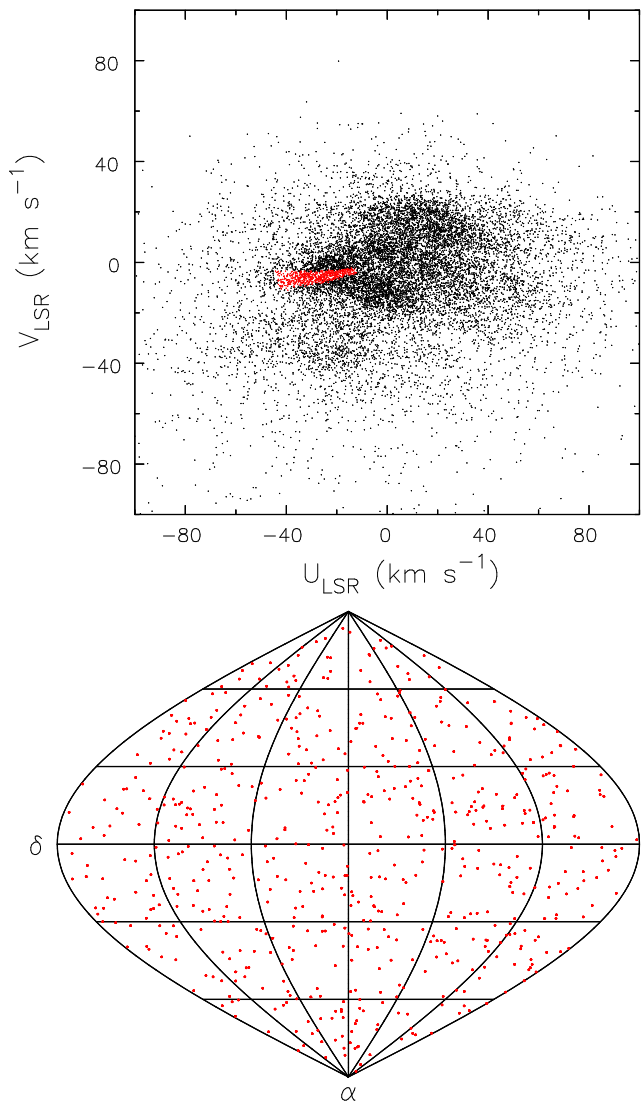}
\caption{The upper panel reproduces the distribution of \GCS\ stars in
  the top-right panel of Fig.~\ref{dtbns}, but with those shown in red
  being the 639 stars picked out in Fig.~\ref{polar}.  The lower panel
  shows the positions, in right ascension and declination, of the
  highlighted stars.}
\label{uvplot}
\end{figure}

\subsection{Properties of the Resonant Stars}
\label{restrs}
The upper panel of Fig.~\ref{uvplot} shows that the 639 stars that I
coloured red in Fig.~\ref{polar} lie in the ``Hyades stream'' in
$(U,V)$-space (feature A of Fig.~\ref{uvcont}).  Note that the
highlighted stars do not include all the resonant stars; thus the fact
that not all the Hyades stream stars are coloured, and that a few
coloured stars are outside the stream may not be significant (see
\S\ref{other}).  It therefore seems clear that the larger part of the
Hyades stream is caused by the \ILR\ identified here.

Aside from their kinematics, I have been unable to find any property
of these 639 stars that clearly differs from the parent distribution
of the \GCS\ sample.  They have a similar distribution of distances,
and incidence of binarity, for example.  Note that \cite{Fama07} do
find a mild, but significant, metallicity excess among the Hyades
stars compared with the mean for the \GCS, although \cite{BH09} find
they are ``barely distinguishable'' from the background population.

The lower panel of Fig.~\ref{uvplot} shows that the 639 red stars are
almost uniformly distributed over the sky.  Note that I have excluded
stars in the Hyades cluster itself from all the analysis because it
would be wrong to treat all the stars within a gravitational bound
cluster as if they were dynamically independent.  Had I included the
cluster, an additional 84 of the 112 Hyades stars would have been
coloured red in Fig.~\ref{polar}, the resonant feature would have
appeared yet more significant, and the cluster would have stood out as
a strong concentration near ${(\alpha,\delta)} = (4^{\rm h}\; 27^{\rm
  m}, 15^\circ 52^\prime)$.  Thus the Hyades cluster itself is resonant.

\subsection{The Age of the Resonance}
Were reliable ages for the stars available, we could use them to date
the origin of the scattering peak, since it can contain stars of all
ages except those that are younger than the wave that caused it.  This
is because the gas from which stars form is not subject to the laws of
collisionless dynamics; a gas cloud in a resonance will experience the
same gravitational accelerations as do the stars, but dissipative
collisions will quickly destroy the resonant signature in phase space.

While age estimates for individual stars appear to be unreliable
\citep{Reid07,Holm07}, we can at least use the Hyades cluster, which
does appear to be resonant (\S\ref{restrs}).  The cluster must have
been formed before the perturbation that scattered it, which places an
upper bound on the time since the resonance of $648\pm45\;$Myr
\citep[\eg][]{dGvH}.

\begin{figure}
\hbox to \hsize{\hss\includegraphics[width=.97\hsize]{uvcolor.ps}}
\smallskip
\includegraphics[width=\hsize]{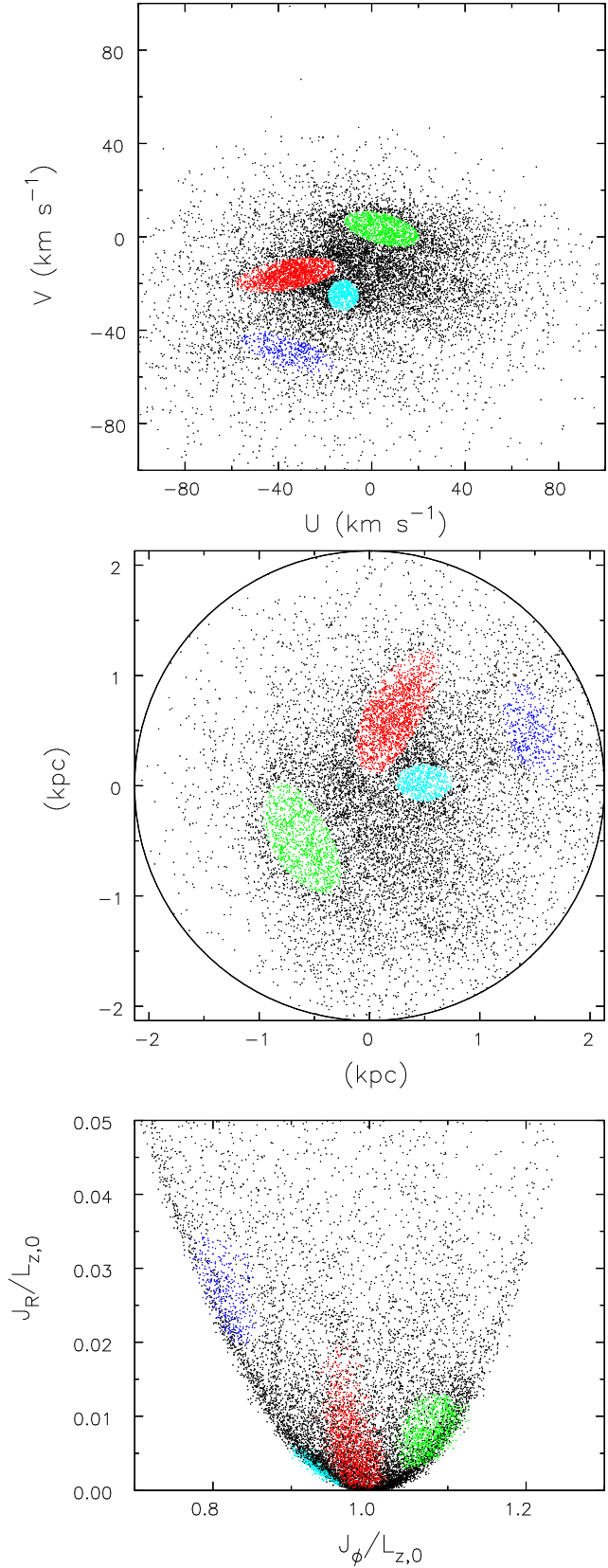}
\caption{The top panel assigns colours to some of the stars near the
  centres of the different streams: Hyades -- red, Sirius -- green,
  Hercules -- dark blue, and Pleiades - cyan.  The same colour codes
  are used to show the locations of these streams in the distribution
  of action and angle variables in the lower two panels.}
\label{streams}
\end{figure}

We can use also use the spread in the spiral phase, or the width of
the peak in Fig.~\ref{restest}, to argue that the perturbation must
have been recent.  Assuming spirals to be transient, the resonance
will have had a width, $\Delta \Omega_p = 0.1\Omega_p$, say.  Thus the
width of the spiral phase will increase with time as
$\Delta(mw_\phi+lw_R) = 0.1m\Omega_p(t-t_0)$.  A generous estimate of
the width of the peak is $0.3$ radians (or $\sim 19^\circ$, over twice
the width of the red triangle in Fig.~\ref{polar}) and therefore the
time since the resonance $t-t_0 < 0.3/(0.1m\Omega_p$), where the
inequality arises because the width can only be increased by
measurement error.  The value of $\Omega_p$ is given by the position
of the peak in Fig.~\ref{linint}, or $\Omega_p = 0.3 V_0/R_0$.  Using
$R_0=8\;$kpc and $V_0=220\;$km~s$^{-1}$ and adopting $m=2$, I find
$t-t_0 \la 180\;$Myr, or the resonance occurred within the last
Galactic rotation.  This rough calculation made a number of
assumptions; assuming a sharper resonance or a broader spread of
spiral phases would increase the time scale, but the values adopted
seem reasonable.  It is reassuring that this estimate is substantially
less than the age of the Hyades.  I therefore conclude that the
resonance is indeed very recent.

Note that not only will phase mixing blur resonances in angle space,
but we must expect resonant features to become less distinct in action
space over time because stars are also scattered by giant molecular
clouds \citep[\eg][]{BL88}.  The rate is hard to quantify, however,
since the rate of cloud scattering is uncertain \citep{Lace91,HF02}.

\subsection{Other ``Streams''}
\label{other}
Since the red stars in Fig.~\ref{polar} turned out to be the Hyades
stream, I have attempted to identify in the same Figure the other
principal features named in Fig.~\ref{uvcont}.  I have coloured all
the \GCS\ stars lying in selected elliptical areas in velocity space
in the top panel of Fig.~\ref{streams}, and the same colours are used
for each star in the lower two panels.  Again I use red for the Hyades
region, but here the selection deliberately errs on generous side to
include most of the stream stars and more besides.

Tracing these stars in other panels reveals that most of the Hyades
stream stars lie in the resonance ridge in Fig.~\ref{actplt} and have
the range of phases that yielded the peak in Fig.~\ref{restest}.  The
Sirius stream (green) appears not to be resonant, as its stars extend
over a significant range of $J_\phi$ but not in $J_R$, and have no
special concentration in phases.  Both the Pleiades (cyan) and
Hercules (blue) stars are near the low-$J_\phi$ boundary of the sample
and may be parts of larger structures that extend outside the sample,
if so their range of phases may be limited.  This selection effect
makes it impossible for this diagnostic to determine whether they are
associated with a resonance.

\subsection{Implications for Spiral Structure Theory}
The identification of an \ILR\ in the solar neighbourhood is a
powerful discriminant between the different theories of spiral pattern
generation.  It has long been known that spiral waves are damped at an
\ILR\ \citep[\eg][p.~502]{BT08}.  \cite{BL96}, who argue that spiral
patterns are long-lived waves, therefore require the \ILR\ to be
shielded in order to avoid fierce damping.  Thus evidence for an
\ILR\ excludes their quasi-steady spiral hypothesis as the originating
mechanism for the disturbance that created it.  While this evidence
does not exclude the possibility of steady spirals elsewhere, it seems
unattractive to invoke a different mechanism to account for spirals in
places other than where we can mount a test.

Strong \ILR\ damping of individual patterns therefore requires spirals
to be continuously regenerated in order that they be common features
in galaxies.  While \cite{TK91} envisage spiral patterns as recurring
transient responses, only my idea \citep{SL89,Sell00} invokes resonant
scattering as the {\it cause\/} of a recurrent cycle of true
instabilities.  Indeed, I predicted \citep{Sell94} that such a feature
might be visible in the \Hipp\ data.

\section{Conclusions}
The foregoing analysis and tests provide compelling evidence that the
``Hyades stream'' in the solar neighbourhood has been created through
scattering at a recent inner Lindblad resonance.  The strongest
evidence is that the stars clustered over a narrow range of the
combination of angles expected for an \ILR\ are also those that lie in
the highly significant overdensity in action space, as indicated by
those coloured red in Fig.~\ref{polar}.  These same stars, which are
distributed all over the sky, make up the Hyades stream as shown in
Fig.~\ref{uvplot}.

\cite{BH09} construct a plot similar to Fig.~\ref{actplt}, also using
the \GCS\ sample but restricted to the $\sim9\,500$ stars within
100~pc of the Sun.  Instead of using radial action as the ordinate,
their Figure~17 uses the almost equivalent energy of non-circular
motion, $E-E_c$, and they scale their plot to include a larger range
of epicycle sizes.  Because of the difference in scale, the feature I
report here is rather inconspicuous in their Figure, but their aim was
to test for the feature I had claimed a hint of in \cite{Sell00}.

Thus only the Hyades stream (feature A of Fig.~\ref{uvcont}) can be
attributed to this \ILR\ and the remaining structures in the local
phase space distribution must be accounted for in other ways.  It is
likely that the \OLR\ of the bar in the Milky Way accounts for the
Hercules stream \citep[feature C,][]{Dehn00}, but we still lack
convincing models to account for the Pleiades and Sirius streams.
Since neither feature stands out in Fig.~\ref{streams} \citep[see also
  Fig.~17 of][]{BH09}, these kinematic groupings of stars have
probably not been caused by resonances with spirals, although they
could perhaps still reflect the continued presence of spirals in this
part of the disc with the resonances lying elsewhere \citep[\eg][]{dSWT}.

Most theoretical studies of spiral structure start with the assumption
that the \DF\ of disk stars is smooth and featureless.  The immediate
implication of a prominent feature in Fig.~\ref{actplt} is that this
assumption is incorrect.  The fact that it was created by an \ILR\ is
strong evidence against the disturbance being a quasi-steady spiral of
the kind proposed by \cite{BL96}.  Rather, it supports the general
picture of spiral structure generation that I have been advocating
\citep[\eg][]{SL89,Sell00}.

Since a spiral pattern should extend outwards from an \ILR, it is
somewhat surprising to find evidence for one this far out in the
Galaxy.  However, the transient spiral waves manifested by simulations
of disc galaxies have a wide spread of pattern speeds
\cite[\eg][]{Sell89}, and do generally include some spirals in the far
outer parts of the disc.  The actual pattern speed implied by this
resonance and the radii of corotation assuming it is a 2-, 3- or
4-armed spiral, are given in Table~\ref{table}; the value for $m=2$ is
clearly lower than other estimates for spirals in the Milky Way
\citep[see][for an up-to-date summary]{Gerh10}.  While the many
assumptions, such as the slope of the local rotation curve, the value
of the \LSR, the neglect of non-axial symmetry, \etc, are potential
sources of large systematic errors in the {\it values\/} given in
Table~\ref{table}, I have argued that none compromise the existence of
the resonance.  If the disturbance responsible for the \ILR\ is a
spiral, then a multi-arm pattern seems more likely than a bi-symmetric
one, but it may also have resulted from some other type of
slowly-rotating disturbance in the outer galaxy, such as a rotating
bi-symmetric distortion of the halo.

Once the still more precise and much more extensive data from {\it
  Gaia\/} \citep{Perr01} become available, it will be possible to
repeat this analysis with far larger samples of stars, and even at
locations other than the solar neighbourhood.  In particular, it
should be possible to follow the azimuthal variation of stars in the
resonance discovered here to determine its angular periodicity.  It is
to be hoped that other resonant scattering features will be found, and
it may be possible to piece together the history of spiral waves in
the Milky Way.

\section*{Acknowledgments}
I thank Scott Tremaine for suggesting that I examine the distribution
of phase angles and Tad Pryor for a number of helpful suggestions on
the statistical analysis.  This work was supported by grants
AST-0507323 from the NSF and NNG05GC29G from NASA.

\label{lastpage}


\begin{thebibliography}{99}

\def\aap{A\&A} \def\aj{AJ} \def\apj{ApJ} \def\apjl{ApJL}
\def\apjs{ApJS} \def\apss{Ap.\ Sp.\ Sci.}  \def\araa{ARAA}
\def\jcoph{J. Comp.\ Phys.}  \def\mnras{MNRAS} \def\PhD{PhD.\ thesis}
\def\nat{Nature} \def\rpp{Rep.\ Prog.\ Phys.}

\bibitem[\protect\citeauthoryear{Antoja \etal}{2009}]{Anto09}
Antoja, T., Valenzuela, O., Pichardo, B., Moreno, E., Figueras, F. \& Fernández, D. 2009, \apjl, {\bf 700}, L78

\bibitem[\protect\citeauthoryear{Aumer \& Binney}{2009}]{AB09}
Aumer, M. \& Binney, J. J. 2009, \mnras, {\bf 397}, 1286

\bibitem[\protect\citeauthoryear{Bensby \etal}{2007}]{Bens07}
Bensby, T., Oey, M. S., Feltzing, S. \& Gustafsson, B. 2007, \apjl, {\bf 655}, L89

\bibitem[\protect\citeauthoryear{Bertin \& Lin}{1996}]{BL96}
Bertin, G. \& Lin, C. C. 1996, {\it Spiral Structure in Galaxies\/} (Cambridge: The MIT Press)

\bibitem[\protect\citeauthoryear{Binney \& Lacey}{1988}]{BL88}
Binney, J. J. \& Lacey, C. G. 1988, \mnras, {\bf 230}, 597

\bibitem[\protect\citeauthoryear{Binney \& Tremaine}{2008}]{BT08}
Binney, J. \& Tremaine, S. 2008, {\it Galactic Dynamics\/} 2nd Ed.\ (Princeton: Princeton University Press), (BT08)

\bibitem[\protect\citeauthoryear{Bovy \& Hogg}{2009}]{BH09}
Bovy, J. \& Hogg, D. W. 2009, arXiv:0912.3262

\bibitem[\protect\citeauthoryear{Bovy \etal}{2009}]{BHR9}
Bovy, J., Hogg, D. W. \& Roweis, S. T. 2009, \apj, {\bf 700}, 1794

\bibitem[\protect\citeauthoryear{Carraro \etal}{2010}]{Carra}
Carraro, G., Vazquez, R. A., Costa, E., Perren, G. \&  Mointinho, A. 2010, arXiv:1006.1277

\bibitem[\protect\citeauthoryear{Chakrabarty}{2007}]{Chak07}
Chakrabarty, D. 2007, \aap, {\bf 467}, 145

\bibitem[\protect\citeauthoryear{De Gennaro \etal}{2009}]{dGvH}
De Gennaro, S., von Hippel, T., Jefferys, W. H., Stein, N., van Dyk, D. \& Jeffery, E. 2009, \apj, {\bf 696}, 12

\bibitem[\protect\citeauthoryear{Dehnen}{1998}]{Dehn98}
Dehnen, W. 1998, \aj, {\bf 115}, 2384

\bibitem[\protect\citeauthoryear{Dehnen}{2000}]{Dehn00}
Dehnen, W. 2000, \aj, {\bf 119}, 800

\bibitem[\protect\citeauthoryear{Dehnen \& Binney}{1998}]{DB98}
Dehnen, W. \& Binney, J. J. 1998, \mnras, {\bf 298}, 387

\bibitem[\protect\citeauthoryear{De Simone \etal}{2004}]{dSWT}
De Simone, R. S., Wu, X. \& Tremaine, S. 2004, \mnras, {\bf 350}, 627

\bibitem[\protect\citeauthoryear{Eggen}{1996}]{Egge96}
Eggen, O. J. 1996, \aj, {\bf 112}, 1595

\bibitem[\protect\citeauthoryear{ESA}{1997}]{ESA7}
ESA 1997, The Hipparcos and Tycho Catalogues, SP 1200

\bibitem[\protect\citeauthoryear{Famaey \etal}{2005}]{Fama05}
Famaey, B., Jorissen, A., Luri, X., Mayor, M., Udry, S., Dejonghe, H. \& Turon, C. 2005, \aap, {\bf 430}, 165

\bibitem[\protect\citeauthoryear{Famaey \etal}{2007}]{Fama07}
Famaey, B., Pont, F., Luri, X., Udry, S., Mayor, M. \& Jorrissen, A. 2007, \aap, {\bf 461}, 957

\bibitem[\protect\citeauthoryear{Gerhard}{2010}]{Gerh10}
Gerhard, O. 2010, arXiv:1003.2489

\bibitem[\protect\citeauthoryear{H\"anninen \& Flynn}{2002}]{HF02}
H\"anninen, J. \& Flynn, C. 2002, \mnras, {\bf 337}, 731

\bibitem[\protect\citeauthoryear{Helmi \etal}{2006}]{Helm06}
Helmi, A., Navarro, J. F., Nordstr\"om, B., Holmberg, J., Abadi, M. G. \& Steinmetz, M. 2006, \mnras, {\bf 365}, 1309

\bibitem[\protect\citeauthoryear{H\o g \etal}{2000}]{Tycho}
H\o g, E., Fabricius, C., Makarov, V. V., Urban, S., Corbin, T., Wycoff, G., Bastian, U., Schwekendiek, P. \& Wicenec, A. 2000, \aap, {\bf 355}, L27

\bibitem[\protect\citeauthoryear{Holmberg \etal}{2007}]{Holm07}
Holmberg, J., Nordstr\"om, B. \& Andersen, J. 2007, \aap, {\bf 475}, 519

\bibitem[\protect\citeauthoryear{Holmberg \etal}{2009}]{HNA9}
Holmberg, J., Nordstr\"om, B. \& Andersen, J. 2009, \aap, {\bf 501}, 941

\bibitem[\protect\citeauthoryear{Kalnajs}{1991}]{Kaln91}
Kalnajs, A. J. 1991, in {\it Dynamics of Disc Galaxies}, ed.\ B. Sundelius (Gothenburg: G\"oteborgs University) p.~323

\bibitem[\protect\citeauthoryear{Lacey}{1991}]{Lace91}
Lacey, C. G. 1991, in {\it Dynamics of Disc Galaxies}, ed.\ B. Sundelius (Gothenburg: G\"oteborgs University) p.~257

\bibitem[\protect\citeauthoryear{Lynden-Bell \& Kalnajs}{1972}]{LBK}
Lynden-Bell, D. \& Kalnajs, A. J. 1972, \mnras, {\bf 157}, 1

\bibitem[\protect\citeauthoryear{Nordstr\"om \etal}{2004}]{GCS}
Nordstr\"om, B., Mayor, M., Andersen, J., Holmberg, J., Pont, F., J\o rgensen, B. R., Olsen, E. H., Udry, S. \& Mowlavi, N.  2004, \aap, {\bf 418}, 989

\bibitem[\protect\citeauthoryear{Perryman \etal}{2001}]{Perr01}
Perryman, M. A. C., de Boer, K. S., Gilmore, G., H\o g, E., Lattanzi, M. G., Lindegren, L., Luri, X., Mignard, F., Pace, O. \& de Zeeuw, P. T. 2001, \aap, {\bf 369}, 339

\bibitem[\protect\citeauthoryear{Quillen}{2003}]{Quil03}
Quillen, A. C. 2003, \aj, {\bf 125}, 785

\bibitem[\protect\citeauthoryear{Quillen \& Minchev}{2005}]{QM05}
Quillen, A. C. \& Minchev, I. 2005, \aj, {\bf 130}, 576

\bibitem[\protect\citeauthoryear{Reid \etal}{2007}]{Reid07}
Reid, I. N., Turner, E. L., Turnbull, M. C., Mountain, M. \& Valenti, J. A. 2007, \apj, {\bf 665}, 767

\bibitem[\protect\citeauthoryear{Robin \etal}{1992}]{RCM92}
Robin, A. C., Cr\'ez\'e, M. \& Mohan, V. 1992, \apjl, {\bf 400}, L25

\bibitem[\protect\citeauthoryear{Sch\"onrich \etal}{2010}]{SBD}
Sch\"onrich, R., Binney, J. \& Dehnen, W. 2010, \mnras, {\bf 403}, 829

\bibitem[\protect\citeauthoryear{Sellwood}{1989}]{Sell89}
Sellwood, J. A. 1989, in {\it Dynamics of Astrophysical Discs}, ed.\ J. A. Sellwood (Cambridge: Cambridge University Press) p.~155

\bibitem[\protect\citeauthoryear{Sellwood}{1994}]{Sell94}
Sellwood, J. A. 1994, in {\it Galactic and Solar System Optical Astrometry} ed.\ L. Morrison (Cambridge: Cambridge Univ.\ Press) p.~156

\bibitem[\protect\citeauthoryear{Sellwood}{2000}]{Sell00}
Sellwood, J. A. 2000, in {\it Astrophysical Dynamics -- in Commemoration of F. D. Kahn}, eds.\ D. Berry, D. Breitschwerdt, A. da Costa \& J. E. Dyson, \apss, {\bf 272}, 31 (astro-ph/9909093)

\bibitem[\protect\citeauthoryear{Sellwood \& Binney}{2002}]{SB02}
Sellwood, J. A. \& Binney, J. J. 2002, \mnras, {\bf 336}, 785

\bibitem[\protect\citeauthoryear{Sellwood \& Lin}{1989}]{SL89}
Sellwood, J. A. \& Lin, D. N. C. 1989, \mnras, {\bf 240}, 991

\bibitem[\protect\citeauthoryear{Silverman}{1986}]{Silv86}
Silverman, B. W. 1986, {\it Density Estimation for Statistics and Data Analysis} (Chapman and Hall, New York)

\bibitem[\protect\citeauthoryear{Toomre \& Kalnajs}{1991}]{TK91}
Toomre, A. \& Kalnajs, A. J. 1991, in {\it Dynamics of Disc Galaxies}, ed.\ B. Sundelius (Gothenburg: G\"oteborgs University) p.~341

\bibitem[\protect\citeauthoryear{van Leeuwen}{2007}]{vL07}
van Leeuwen, F. 2007, \aap, {\bf 474}, 653

\end{thebibliography}
\end{document}